\documentclass[twocolumn]{aastex63}

\shorttitle{Dynamical Packing in the Habitable Zone}
\shortauthors{Stephen R. Kane et al.}

\begin{document}

\title{Dynamical Packing in the Habitable Zone: The Case of Beta CVn}

\author[0000-0002-7084-0529]{Stephen R. Kane}
\affiliation{Department of Earth and Planetary Sciences, University of
  California, Riverside, CA 92521, USA}
\email{skane@ucr.edu}

\author[0000-0002-0569-1643]{Margaret C. Turnbull}
\affiliation{SETI Institute, Carl Sagan Center for the Study of Life
  in the Universe, Off-Site: 2613 Waunona Way, Madison, WI 53713, USA}

\author[0000-0003-3504-5316]{Benjamin J. Fulton}
\affiliation{Caltech/IPAC-NASA Exoplanet Science Institute, Pasadena,
  CA 91125, USA}

\author[0000-0001-8391-5182]{Lee J. Rosenthal}
\affiliation{Department of Astronomy, California Institute of
  Technology, Pasadena, CA 91125, USA}

\author[0000-0001-8638-0320]{Andrew W. Howard}
\affiliation{Department of Astronomy, California Institute of
  Technology, Pasadena, CA 91125, USA}

\author[0000-0002-0531-1073]{Howard Isaacson}
\affiliation{Department of Astronomy, University of California,
  Berkeley, CA 94720, USA}
\affiliation{Centre for Astrophysics, University of Southern
  Queensland, Toowoomba, QLD 4350, Australia}

\author[0000-0002-2909-0113]{Geoffrey W. Marcy}
\affiliation{Department of Astronomy, University of California,
  Berkeley, CA 94720, USA}

\author[0000-0002-3725-3058]{Lauren M. Weiss}
\affiliation{Institute for Astronomy, University of Hawai`i, Honolulu,
  HI 96822, USA}


\begin{abstract}

Uncovering the occurrence rate of terrestrial planets within the
Habitable Zone (HZ) of their host stars has been a particular focus of
exoplanetary science in recent years. The statistics of these
occurrence rates have largely been derived from transiting planet
discoveries, and have uncovered numerous HZ planets in compact systems
around M dwarf host stars. Here we explore the width of the HZ as a
function of spectral type, and the dynamical constraints on the number
of stable orbits within the HZ for a given star. We show that,
although the Hill radius for a given planetary mass increases with
larger semi-major axis, the width of the HZ for earlier-type stars
allows for more terrestrial planets in the HZ than late-type stars. In
general, dynamical constraints allow $\sim$6 HZ Earth-mass planets for
stellar masses $\gtrsim 0.7 M_\odot$, depending on the presence of
farther out giant planets. As an example, we consider the case of Beta
CVn, a nearby bright solar-type star. We present 20 years of radial
velocities (RV) from the Keck/HIRES and APF instruments and conduct an
injection-recovery analysis of planetary signatures in the data. Our
analysis of these RV data rule out planets more massive than Saturn
within 10~AU of the star. These system properties are used to
calculate the potential dynamical packing of terrestrial planets in
the HZ and show that such nearby stellar targets could be particularly
lucrative for HZ planet detection by direct imaging exoplanet
missions.

\end{abstract}



\section{Introduction}
\label{intro}

The common exoplanet detection techniques of transits and radial
velocities (RV) both have a bias towards short-period orbits
\citep{kane2008b}. The advantage of this bias is that it has enabled a
thorough exploration of multiple planet systems that exists in closely
spaced orbits, often in extreme insolation flux environments, referred
to as ``compact systems''. The architecture and dynamics of compact
planetary systems have been previously studied in detail
\citep{ford2014}, including the dynamical connection to formation
\citep{kane2013e,pu2015} and the exclusion of moons
\citep{kane2017c}. \citet{barnes2004a} suggested that planetary
systems may have quantifiable limits to the dynamically allowed number
of planets. Simulations by \citet{smith2009,smith2010} and
\citet{obertas2017} investigated dynamical interactions within compact
systems for the specific solar case and found a dependence on planet
mass and initial spacing. A statistical analysis of compact systems by
\citet{fang2013} indicated that compact systems are dynamically
``packed'', with few dynamically viable options for additional planets
in these systems. Furthermore, the presence of farther planetary
companions in compact systems can be effectively constrained due to
their disrupting influence on the inner planets \citep{becker2017a}.

Of particular interest are those planets that lie within the Habitable
Zone (HZ) of their host stars, where liquid water may be present on
the planetary surface given sufficient atmospheric pressure
\citep{kasting1993a,kopparapu2013a,kopparapu2014}. Dynamics of
planetary systems have been used to determine orbital effects in the
HZ \citep{kane2015b,agnew2019}, as well as potentially predicting
additional HZ planets \citep{kopparapu2009,kopparapu2010}. Several
compact planetary systems have been discovered that harbor more than
one planet in the HZ, including GJ~667C \citep{angladaescude2013c},
TRAPPIST-1 \citep{gillon2017a,luger2017b}, and Teegarden's star
\citep{zechmeister2019}. Compact HZ systems around M dwarfs may
benefit from numerous architectural biases for low-mass stars. For
example, it has been suggested from observational data that giant
planets are relatively rare around M dwarfs
\citep{endl2006b,cumming2008,johnson2010d,bonfils2013a} and that this
may influence the habitability of terrestrial planets in those systems
\citep{horner2008a,shields2016b,kane2019e}. The inventory of confirmed
exoplanets provided by the NASA Exoplanet Archive \citep{akeson2013},
extracted 2020 May 11, indicates that planets more massive than 0.5
Jupiter masses orbiting stars less massive than 0.5 solar masses
comprise less than 2\% of the total RV exoplanet discoveries. The
architecture, exoplanet detection sensitivity, and prevalence of HZ
planets in M dwarf systems clearly play significant roles in
calculations of occurrence rates of HZ terrestrial planets
\citep{catanzarite2011a,dressing2013,gaidos2013b,kopparapu2013b,foremanmackey2014,dressing2015b}. The
dynamics and prevalence of HZ terrestrial planets have yet to be as
rigorously tested around solar-type stars.

A star of particular interest is the nearby (8.44~pcs) star Beta Canum
Venaticorum. Being a naked eye star ($V = 4.26$), the star has various
designations (HD~109358, HIP~61317) including the name ``Chara'', but
is hereafter referred to as Beta CVn. The star has been identified as
a target of astrobiological significance due to both its similarity
and proximity to the Sun
\citep{portodemello2006,turnbull2015}. Although no planets have been
announced for the star, the limits on planetary companions based on
survey data have yet to be demonstrated. Such limits on giant planets
would potentially allow for greater dynamical diversity within the HZ,
including scenarios of dynamically packed orbits that may otherwise be
excluded due to secular resonances with giant planets in the system
\citep{levison2003a,brasser2009,kopparapu2010}.

This paper explores the dynamical packing of terrestrial planets in
the HZ as a function of spectral type. Section~\ref{hz} uses stellar
isochrones to quantify the width of the HZ as a function of stellar
mass, describes the results of dynamical simulations of HZ packing of
terrestrial planets, and demonstrates the effect of giant planets on
HZ stability. In Section~\ref{betacvn} we present the results of the
Beta~CVn case study, including stellar characterization, HZ region,
and 20 years RV data from the Keck/HIRES and Automated Planet Finder
(APF) facilities that rule out giant planets in the system. These data
are used to estimate the potential for terrestrial planets in the HZ
and prospects for direct imaging observations. We provide discussion
of applications to future work and general conclusions in
Section~\ref{conclusions}.


\section{Habitable Zone Real Estate}
\label{hz}


\subsection{Stellar Mass Dependence}
\label{smass}

The boundaries of the HZ have been cataloged for a variety of known
exoplanetary systems \citep{kane2012a}, including the candidate
exoplanet systems discovered by the {\it Kepler} mission
\citep{kane2016c,hill2018}. The stellar distance plays a key role in
determining the stellar properties that influence the HZ boundaries,
highlighting the importance of accurate distance estimates
\citep{johns2018,kane2018a}. Here we calculate the width of the HZ as
a function of the stellar mass, since mass is one of the more critical
intrinsic stellar properties that determines the overall stellar
evolution pathway.

\begin{figure}
  \includegraphics[angle=270,width=8.5cm]{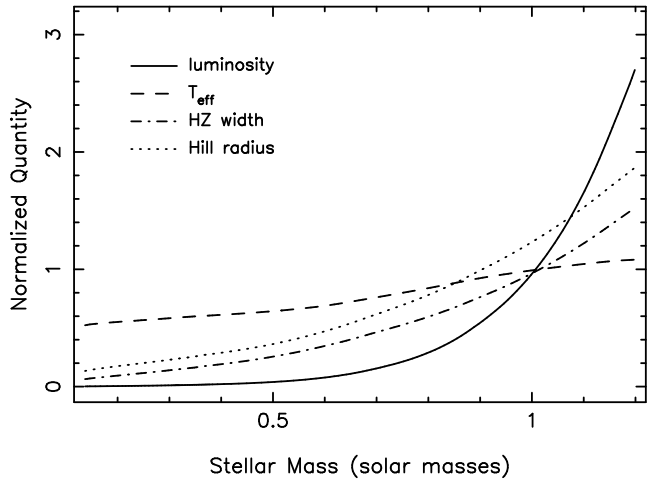}
  \caption{Stellar luminosity (solid line) and effective temperature
    (dashed line) as a function of stellar mass. For each of these
    values, we calculate the HZ width (dot-dashed line) and the Hill
    radius of an Earth-mass planet in the middle of the HZ (dotted
    line). For comparative purposes, the luminosity, effective
    temperature, and HZ have been normalized to solar values, and the
    Hill radius is normalized to the Earth value.}
  \label{hzwidth}
\end{figure}

To investigate the width of the HZ region as a function of stellar
mass for main sequence stars, we utilize the MESA Isochrones \&
Stellar Tracks (MIST), described in detail by \citet{choi2016} and
\citet{dotter2016}. We adopted a solar metallicity and used an age of
3~Gyrs in order to encompass isochrones for stars more massive than
the Sun, including a mass range of 0.137--1.199~$M_\odot$. For each
stellar mass, we used the isochrone luminosity and effective
temperature to calculate the HZ boundaries as formulated by
\citet{kopparapu2013a,kopparapu2014}. We adopt both the
``conservative'' and ``optimistic'' HZ boundaries, described in detail
by \citet{kane2016c}.

\begin{figure*}
  \begin{center}
    \includegraphics[angle=270,width=16.0cm]{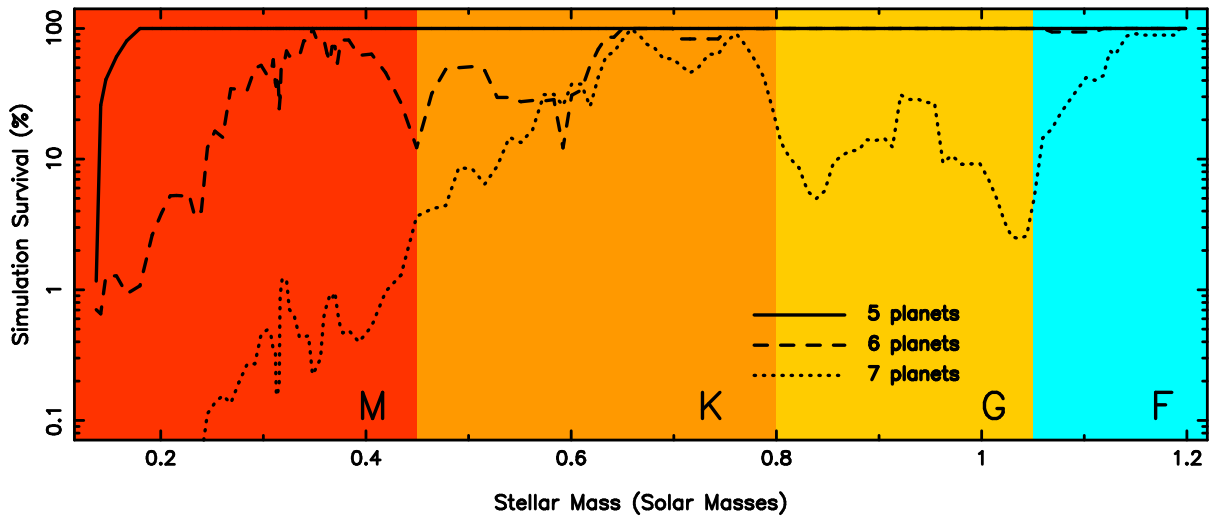}
  \end{center}
  \caption{Simulation survival of the tested planetary architecture as
    a function of stellar mass, where the architecture consisted of
    evenly spaced Earth-mass planets for 3, 4, 5, 6, and 7 planets in
    the HZ, as described in Section~\ref{packing}. The cases of 5, 6,
    and 7 planets are shown as solid, dashed line, dotted lines
    respectively. The colored backgrounds indicate the corresponding
    spectral types (see bottom right of each colored region) for the
    stellar mass ranges shown.}
  \label{stability}
\end{figure*}

The extracted isochrone stellar parameters of luminosity (solid line)
and effective temperature (dashed line) along with our calculations of
the HZ width (dot-dashed line) are represented in
Figure~\ref{hzwidth}. The values for all three have been normalized to
solar values, which is why they intersect at a normalized quantity of
unity for a solar mass star. We further calculated the Hill radius for
an Earth-mass planet located in the middle of the HZ for each stellar
mass. These Hill radii were also normalized, this time to the Hill
radius of the Earth, and are represented in Figure~\ref{hzwidth} as a
dotted line. The rise in Hill radius with stellar mass is slightly
less than the rise in HZ width. This implies that dynamical
constraints on the number of planets in the HZ may increase with
increasing stellar mass.


\subsection{Dynamical Packing Limits}
\label{packing}

To test the dynamical limits of terrestrial planets within the HZ, we
conducted an extensive suite of N-body integrations using the Mercury
Integrator Package, described in detail by \citet{chambers1999}. The
simulations adopted the hybrid symplectic/Bulirsch-Stoer integrator
with a Jacobi coordinate system, generally providing more accurate
results for multi-planet systems \citep{wisdom1991,wisdom2006b}. For
each of the stellar masses extracted from the isochrones described in
Section~\ref{smass}, we performed the following steps.
\begin{enumerate}
\item The boundaries of the optimistic HZ were calculated based on the
  stellar parameters.
\item Earth-mass planets were placed in circular orbits within the HZ,
  with one at the inner edge, one at the outer edge, and the others
  evenly spaced in between \citep{weiss2018a}. This was carried out
  for a range of 3--7 planets for each stellar mass.
\item The orbital period at the inner and outer edges of the HZ were
  calculated. The orbital period at the inner edge was used to
  determine the time step of the N-body integrations. These were set
  to $1/20$ of the inner edge orbital period to ensure perturbative
  reliability of the simulation, consistent with the recommendations
  of \citet{duncan1998}. The total simulation time was set to $10^8$
  times the orbital period at the outer edge of the HZ.
\item The dynamical simulation was carried out $\sim$5 times with
  randomized starting locations (mean anomalies) for each of the
  planets to robustly capture the dynamical stability of the orbital
  architecture.
\item The assessment of stability was determined by requiring that all
  planets survive the duration of the simulation. Non-survival means
  that one or more of the planets has been captured by the
  gravitational well of the host star or ejected from the system.
\end{enumerate}
Following the above steps resulted in several thousand simulations to
fully explore the dynamical stability limits of packing evenly spaced
terrestrial planets within the HZ for a range of stellar masses.

The results of the simulation suite described above are shown in
Figure~\ref{stability}, where the colored shaded regions indicate the
approximate spectral types that correspond to the stellar masses. For
the 3 and 4 planet cases, we found that the dynamical integrity of the
systems are retained for the entire range of stellar masses tested. We
show the 5 planet case (solid line) as a demonstration of where
instability begins to arise at the extreme low mass end of the stellar
mass range. The 6 planet case (dashed line) exhibits significant
instability within the HZ for stellar masses less than
$\sim$0.7~$M_\odot$. The trend of relative instability for lower mass
stars is consistent with the relationship between the HZ width and the
size of the Hill sphere discussed in Section~\ref{smass}. Significant
collapse of the dynamical integrity for all stellar masses starts to
occur for the 7 planet case (dotted line), with no stable scenarios
for masses less than $\sim$0.25~$M_\odot$. Furthermore, the stellar
mass range of 0.8--1.1~$M_\odot$ (G stars) is unlikely to harbor 7
evenly spaced planets within the HZ. Calculation of the corresponding
orbital periods of the planets for this stellar mass range reveals a
cause of mean motion resonances (MMRs). Specifically, the third planet
in the sequence of 7 planets (from inner HZ edge to outer HZ edge)
lies close to the 5:4 MMR with planet 4, the 3:2 MMR with planet 5,
and 2:1 MMR with planet 7. These MMRs result in strong perturbative
forces on planet 3 that are able to successfully compromise the system
stability and rapidly lead to close encounters and subsequent
ejections. Thus, in some cases, MMRs can lead to a potential
additional advantage of K dwarfs over G dwarfs for biosignature and
direct imaging studies \citep{arney2019}.

For example, the case of our solar system has an optimistic HZ that
extends from 0.74~AU to 1.73~AU. Our simulations show that 6
terrestrial planets are able to remain stable in this region but the 7
HZ planet architecture falls within the regime of MMR perturbations
and catastrophic instability. This is consistent with the results of
\citet{smith2009} and \citet{obertas2017} whose simulations were for
the specific solar mass case. These simulations neglect the effect of
giant planets beyond the HZ that can also induce perturbative effects
that can disrupt even the 6 planet scenarios.


\subsection{Effects of Giant Planets}
\label{giant}

The number of known exoplanetary systems is now sufficient to develop
statistical models that can predict orbital distributions based on
collisional formation scenarios \citep{tremaine2015}. However, these
distributions are greatly influenced by giant planets, particularly
during their early periods of migration \citep{morbidelli2016b}.  The
presence of the giant planets in the solar system has undoubtedly
played a significant role in the formation, evolution, and final
architecture of the inner solar system
\citep{tsiganis2005b,batygin2015b,horner2020a}. Such important
dynamical effects of giant planets have also shaped the architecture
of other planetary systems \citep{matsumura2013,morbidelli2016b}.
Indeed, both the solar system and other planetary systems have been
found to be close to instability boundaries that would compromise the
dynamical integrity of those systems
\citep{laskar1996b,lecar2001,murray2001a,barnes2004a}. For example, a
planet within the asteroid belt of the solar system would
significantly alter the orbital dynamics of the inner solar system by
exchanging angular momentum with the outer solar system
\citep{lissauer2001c}.

With this in mind, the HZ packing scenarios described in
Section~\ref{packing} have been considered in isolation but are
clearly dependent on the presence of other planets outside of the HZ,
especially giant planets. Jupiter has a well-known perturbing
influence at resonance locations, especially the induced chaos near
the Kirkwood gaps \citep{wisdom1983a}. The case of 7 planets within
the HZ, described in Section~\ref{packing}, suffers from instability
due to MMR effects internal to that region. External sources of MMR,
particularly 5:2, 3:1, would result in similar sources of instability
which would have a cascading transfer of angular momentum, similar to
the asteroid belt planet scenario described by
\citet{lissauer2001c}. Therefore, the most likely cases of maximum
terrestrial planets within the HZ are those solar-type stars for which
giant planets may be excluded.


\section{The Case of Beta CVn}
\label{betacvn}

Here we consider the specific case for potential dynamical HZ packing
of Beta CVn, a nearby solar analog (G0V).


\subsection{Stellar Properties and Habitable Zone}
\label{stellar}

The proximity, spectral type, and apparent brightness of Beta CVn have
resulted in numerous observational studies of the star. The star has
been found to be photometrically and spectroscopically stable,
resulting in its frequent use as an RV standard
\citep{konacki2005b}. A study of Sun-like stars by \citet{radick2018}
presented 18 years of Beta CVn photometry, acquired with the Automated
Photoelectric Telescopes \citep{henry1999}. We extracted the Beta CVn
photometry from their data collection and conducted a Fourier analysis
of the complete time series. The highest peak in the power spectrum
from this analysis occurs at 182.5 days, and is an observational
artifact from the observing cadence. The second highest peak occurs at
110 days, which could be related to intrinsic stellar variability,
although is too long for an expected rotation period.

Given the quiet nature of the star, the question arises as to the
stellar equatorial plane relative to the plane of the sky since this
will influence the detectability of planetary signatures, assuming
those orbits are coplanar with the stellar equator. In order to
optimize blind transit searches, \citet{herrero2012} provided a
catalog of stars whose inclination of the equatorial plane is likely
$> 80\degr$ relative to the plane of the sky. They adopt a rotational
velocity of $v \sin i = 2.9$~km\,s$^{-1}$ and activity index of
$\log(R'_{HK}) = -4.885$ for Beta CVn, and their methodology predicts
a high probability (50\% larger than the average of their sample) of
the star having an inclination $> 80\degr$. The predicted probability
is 10\% higher than the mean probability for the host stars in their
sample that are known to host transiting planets. This in turn
increases the probability that any planets present are in close to
edge-on orbits, increasing the expected RV signal.

The proximity of Beta CVn has also enabled several direct measurements
of the stellar radius using interferometric techniques
\citep{vanbelle2009a,boyajian2012a}. We adopt the stellar parameters
provided by \citet{boyajian2012a}, which include a stellar radius of
$R_\star = 1.123\pm0.028$~$R_\odot$, a luminosity of $L_\star =
1.151\pm0.018$~$L_\odot$ and an effective temperature of
$T_\mathrm{eff} = 5653\pm72$~K. Based on these stellar parameters, we
calculate the conservative HZ as 1.03--1.82~AU and the optimistic HZ
as 0.81--1.92~AU. The significance of the HZ distances for the star
are described in more detail in Section~\ref{rocky}.


\subsection{No Giant Planets}
\label{rvs}

\begin{figure}
  \includegraphics[clip,width=8.5cm]{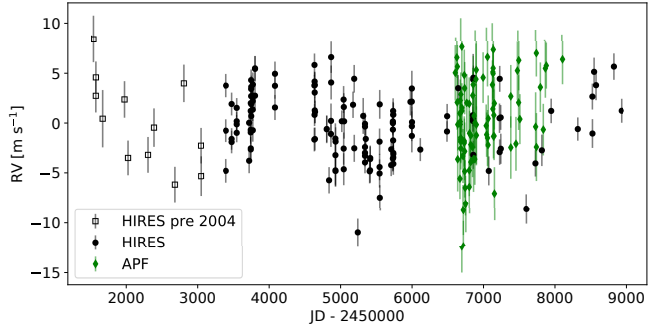}
  \caption{The RV data for Beta CVn plotted against observation time,
    including data from HIRES pre-2004 upgrade (HIRES$_k$), HIRES
    post-2004 upgrade (HIRES$_j$), and APF.}
  \label{velsfig}
\end{figure}

Even though giant planets are more common around solar analogs than
later-type stars \citep{zechmeister2013,wittenmyer2016c}, they are
still relatively rare, with an average occurrence rate of 6.7\% beyond
orbital periods of 300 days \citep{wittenmyer2020a}. Consequently, the
solar system may not be representative of orbital architectures, and
it is useful to directly compare with another nearby solar-type star,
such as Beta CVn. To determine the limits on such giant companions, we
present here 235 precision RV measurements of Beta CVn acquired over a
period of 20 years. Observations were carried out using the HIRES
echelle spectrograph on the Keck I telescope \citep{vogt1994} and the
Levy spectrometer on the Automated Planet Finder (APF)
\citep{radovan2014,vogt2014a}. A subset of the RV data are shown in
Table~\ref{vels}, where all the data have been calibrated to the same
zero-point. Table~\ref{vels} also includes a column for the instrument
that was used to acquire the data. Note that the HIRES subscripts
refer to data that were acquired prior to the 2004 upgrade to the
instrument (k) and those that were acquired after (j). The data from
all three instruments are shown plotted in Figure~\ref{velsfig}.

\begin{deluxetable}{lccc}
  \tablewidth{0pc}
  \tablecaption{\label{vels}Beta CVn Radial Velocities}
  \tablehead{
    \colhead{Instrument} &
    \colhead{Date} &
    \colhead{RV} &
    \colhead{$\sigma$} \\
    \colhead{} &
    \colhead{(BJD -- 2450000)} &
    \colhead{(m\,s$^{-1}$)} &
    \colhead{(m\,s$^{-1}$)}
  }
  \startdata
HIRES$_k$ & 1551.1784 & 0.339 & 2.329\\
HIRES$_k$ & 1583.9661 & -5.363 & 1.669\\
HIRES$_k$ & 1585.0081 & -3.517 & 1.596\\
HIRES$_k$ & 1678.9005 & -7.658 & 2.871\\
HIRES$_k$ & 1982.0561 & -5.728 & 1.830\\
HIRES$_k$ & 2030.8898 & -11.587 & 1.739\\
HIRES$_k$ & 2308.1046 & -11.281 & 1.792\\
HIRES$_k$ & 2391.0513 & -8.550 & 1.926\\
HIRES$_k$ & 2681.0914 & -14.265 & 1.789\\
HIRES$_k$ & 2805.8423 & -4.116 & 1.879\\
HIRES$_k$ & 3044.1802 & -10.360 & 1.780\\
HIRES$_k$ & 3046.1807 & -13.416 & 1.988\\
HIRES$_j$ & 3399.0412 & 4.337 & 1.230\\
HIRES$_j$ & 3399.0418 & -1.512 & 1.197\\
HIRES$_j$ & 3399.0424 & -4.692 & 1.123\\
HIRES$_j$ & 3480.7773 & -1.325 & 1.089\\
HIRES$_j$ & 3480.7779 & 0.645 & 1.100\\
HIRES$_j$ & 3480.7785 & -1.011 & 1.127\\
HIRES$_j$ & 3551.8156 & -0.572 & 1.022\\
HIRES$_j$ & 3551.8162 & -0.750 & 0.918\\
HIRES$_j$ & 3551.8168 & -0.070 & 0.946\\
HIRES$_j$ & 3551.8174 & 0.737 & 1.005\\
HIRES$_j$ & 3725.1611 & -2.464 & 1.253\\
HIRES$_j$ & 3725.1617 & -0.609 & 1.169\\
HIRES$_j$ & 3747.1081 & 0.014 & 1.121
  \enddata
\tablenotetext{}{The full data set is available online.}
\end{deluxetable}

\begin{figure}
  \includegraphics[clip,width=8.5cm]{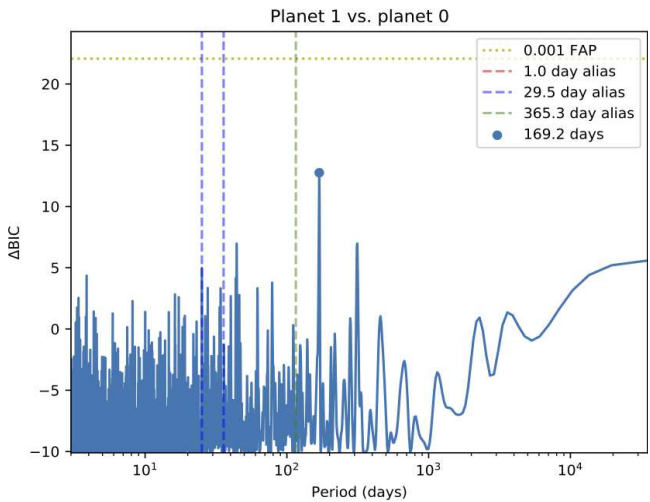}
  \caption{Power spectrum from a Fourier analysis of the RV data shown
    in Figure~\ref{velsfig}. Significant observational alias locations
    (diurnal, lunar, annual) are shown are vertical dashed lines. The
    highest peak occurs at a period of 169.2~days.}
  \label{pgramfig}
\end{figure}

We performed a Fourier analysis of the RV data to investigate possible
periodic signals. The periodogram resulting from this analysis is
shown in Figure~\ref{pgramfig}. The horizontal dotted line indicates a
false-alarm probability (FAP) threshold of 0.001 (0.1\%). The vertical
dashed lines show the location of common aliases caused by the Earth's
rotation (diurnal) and orbital (annual) motions, as well as that
caused by the lunar orbit. The highest peak occurs at a period of
169.2~days, which is still well below the FAP threshold. We conclude
that there are no significant periodic signals present in the data.

The RV data described here were used to perform an injection-recovery
test to quantify the completeness of the observations for excluding
planetary signatures. This method injects planetary signatures of
various masses ($M_p \sin i$) and semi-major axes ($a$) into the data
using the same observation epochs and noise properties of the
data. Circular orbits are assumed and the Keplerian orbital fits are
performed using the RadVel package \citep{fulton2018a}. The
methodology for this process, including the criteria for planetary
signature recovery, is described in detail by \citet{howard2016}. To
perform the injection-recovery test in terms of $M_p \sin i$, a
stellar mass must be assumed. There are a variety of estimates for the
mass of Beta CVn in the literature, including $1.05\pm0.14$~$M_\odot$
by \citet{valenti2005} and $0.852\pm0.023$~$M_\odot$ by
\citet{boyajian2012a}. We adopt the latter of these to be consistent
with the stellar parameters used to calculate the HZ in
Section~\ref{stellar}.

\begin{figure}
  \includegraphics[clip,width=8.5cm]{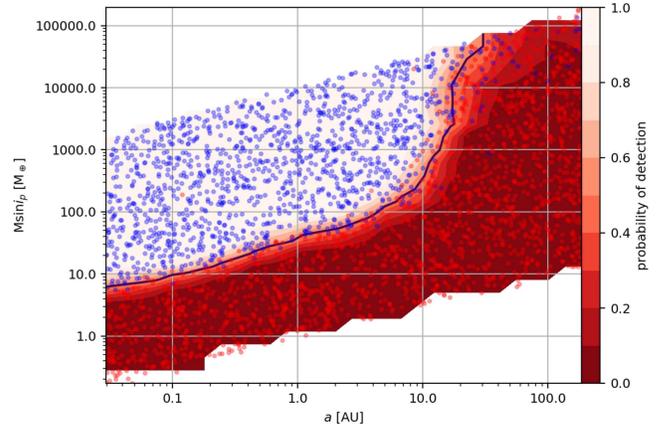}
  \caption{Results of the injection-recovery test to determine the
    sensitivity of the RV data to planetary signatures as a function
    of planetary mass ($M_p \sin i$) and semi-major axis ($a$). The
    blue dots represent injected planetary signatures that were
    successfully recovered and the red dots represent those planets
    that were not recovered. The color scale corresponding to the
    probability contours of detecting a planet of a given mass and
    semi-major axis is shown on the right vertical axis.}
  \label{sensitivity}
\end{figure}

The results of the injection-recovery test are shown in
Figure~\ref{sensitivity}. The blue dots represent injected planets
that were recovered and the red dots represent those that were not
recovered. The shaded contours indicate the probability of a
successful detection, with the color scale shown on the right vertical
axis. The results thus demonstrate that the RV data are sufficient to
rule out the presence of 100~$M_\oplus$ planets out to $\sim$5~AU and
300~$M_\oplus$ (Jupiter mass) planets out to $\sim$10~AU. These data
are thus sufficient to rule out the presence of most giant planets
around Beta CVn inside an orbital radius of 10~AU. Note however that
the data are not sufficient to rule out terrestrial planets in the
system.


\subsection{Potential for Terrestrial Planets}
\label{rocky}

As described in Sections~\ref{packing} and \ref{giant}, the relative
lack of giant planets in the Beta CVn system may allow for the
presence of numerous terrestrial planets in a HZ dynamical packing
scenario. Based on Figure~\ref{stability} and the stellar mass adopted
in Section~\ref{rvs}, the HZ of the star would be unlikely to contain
7 terrestrial planets even without a giant planet due to internal MMR
perturbation effects. Beta CVn may therefore be a suitable system for
which to test HZ packing scenarios.

As calculated in Section~\ref{stellar}, the optimistic HZ of the
system extends from 0.81~AU to 1.92~AU. Continuing to use the stellar
parameters of \citet{boyajian2012a}, we calculate several predicted
exoplanet signatures at the inner and outer edge of the optimistic HZ,
shown in Table~\ref{signatures}. These include the orbital period $P$,
the RV semi-amplitude $K$, the transit depth $d$, and the transit
probability $p$ \citep{kane2008b}. As expected, the $K$ values are
well below that required of the RV data presented in this work, but
may be feasible with developing facilities
\citep{fischer2016}. Additionally, the work of \citet{herrero2012}
showed that Beta CVn has a relatively high probability that the
inclination of the equatorial plane is $> 80\degr$ relative to the
plane of the sky (see Section~\ref{stellar}), and so the transit
probabilities are likely underestimated.

\begin{deluxetable}{lrr}
  \tablecolumns{3}
  \tablewidth{0pc}
  \tablecaption{\label{signatures}HZ Exoplanet Signatures}
  \tablehead{
    \colhead{Parameter} & 
    \colhead{Inner HZ} &
    \colhead{Outer HZ}
  }
  \startdata
$P$ (days)        & 288.4 & 1052.6 \\
$K$ (m\,s$^{-1}$) & 0.108 & 0.070 \\
$d$ (\%)          & 0.007 & 0.007 \\
$p$ (\%)          & 0.65  & 0.27
  \enddata
\end{deluxetable}

However, a further option to explore the possible planetary
architecture of Beta CVn, particularly given the proximity of the
star, is direct imaging \citep{kane2018c,dulz2020}. At a stellar
distance of 8.44~pcs, the 0.81--1.92~AU optimistic HZ translates into
angular separations of 96--227~mas. Consistent with the final reports
of the Habitable Exoplanet Observatory (HabEx) mission
\citep{reporthabex} and the Large UV/Optical/Infrared Surveyor
(LUVOIR) mission \citep{reportluvoir}, we assume (a) an inner working
angle (IWA) of $\sim$58~mas, and (b) a contrast limit of
$\sim$$4\times10^{-11}$, which corresponds to the Earth--Sun contrast
ratio at the outer edge of the Sun's conservative HZ (1.7~AU).

\begin{figure}
  \includegraphics[clip,width=8.5cm]{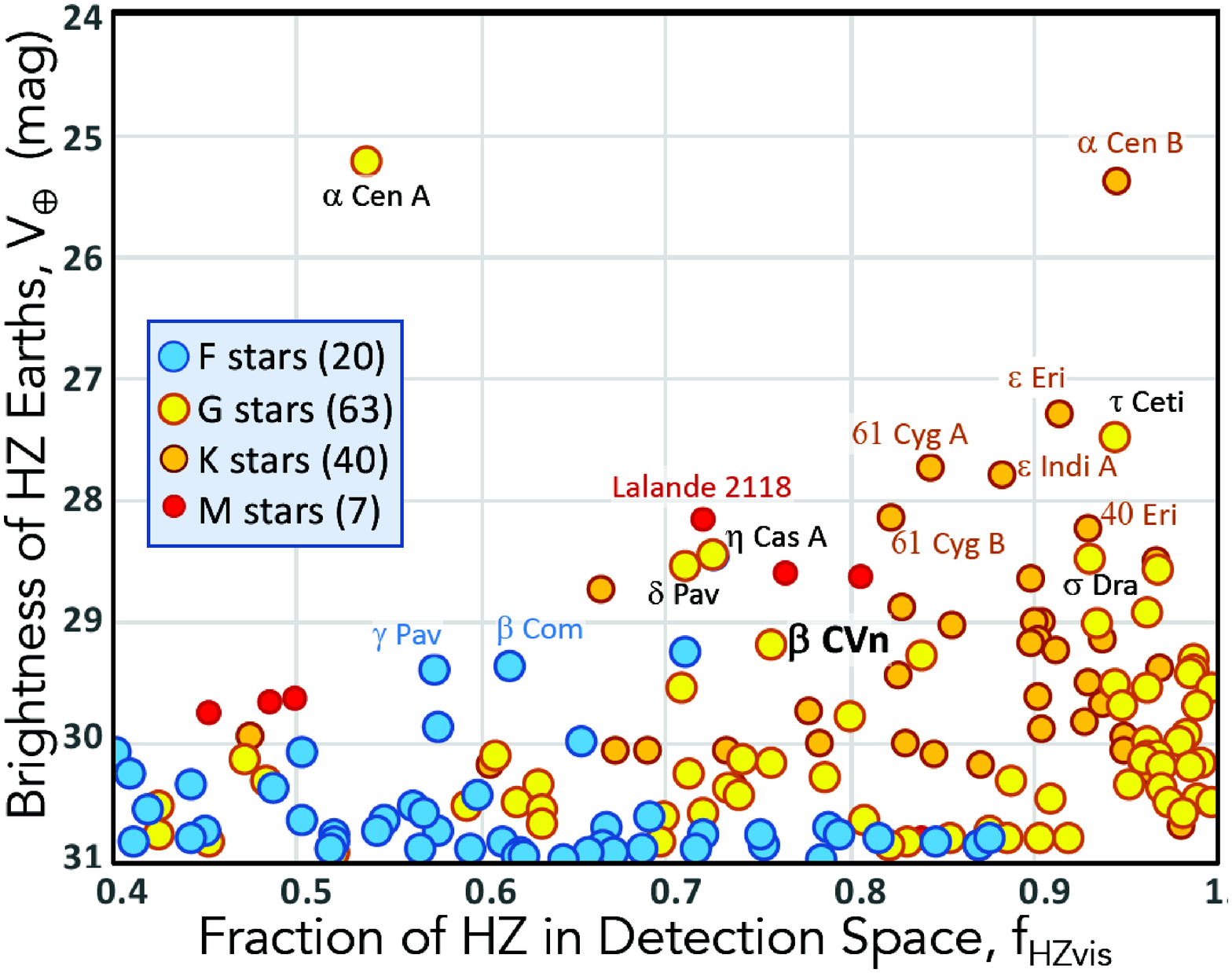}
  \caption{Calculated V-band magnitude of an Earth analog in the HZ of
    the nearest stars, as a function of the fraction of the HZ for
    those stars that are within instrument detection limits, scaled by
    the Hill radius. The color of the stars indicates their effective
    temperature (spectral type) and labels are provided for well-known
    nearby stars, including Beta CVn.}
  \label{imaging}
\end{figure}

Given the above assumptions, Figure~\ref{imaging} shows the stars
having the largest fraction of their HZs falling within the detection
limits of the instrument (scaled by the Hill radius), and the maximum
V-band brightness of Earth analogs within the detectable HZ space,
assuming a Bond albedo of 0.3. The color of the stars shown in the
figure represents the effective temperature (spectral type) of the
host stars, with M red, K orange, G yellow, and F cyan; similar to
Figure~\ref{stability}. The most promising targets are shown in the
right of the plot where both the predicted HZ Earth analog brightness
and the visibility of the HZ are largest. For example, Beta CVn has a
fractional HZ visibility of $\sim$75\% and an expected Earth analog
brightness of $V \sim 29$. Note that the stellar sample shown in
Figure~\ref{imaging} is dominated by K and G dwarfs for which the
occupancy of the HZ by such Earth analogs is maximized, except in
cases where internal MMRs truncate the number of allowed orbits (see
Figure~\ref{stability}). Thus, direct imaging with facilities that
meet the criteria laid down in the HabEx and LUVOIR reports may
naturally favor targets that are predisposed towards larger than
expected HZ planet inventories.


\section{Conclusions}
\label{conclusions}

The search for potentially habitable exoplanets has an intrinsic
relationship to their occurrence rates, a factor that has rapidly
emerged in recent years. These occurrence rates are mostly being
informed by transit surveys that are biased toward HZ planets around
low-mass stars and have revealed a plethora of compact system
architectures. Some of these systems, such as TRAPPIST-1, have been
found to have multiple terrestrial planets in the HZ, causing
speculation as to whether such dynamical packing in the HZ is normal
or rare. Our analysis presented here demonstrates that in fact the
dynamical limitations to the packing of HZ terrestrial planets is
$\sim$5 planets for most spectral types, and $\sim$6 planets for
stellar masses $\gtrsim 0.7 M_\odot$. Packing 7 planets in the HZ is
possible within certain specific stellar mass and architecture
regimes, but becomes vulnerable to MMR perturbations that compromise
the dynamical stability of such configurations. The 20 years of RV
data for Beta CVn presented here rule out a large range of giant
planet masses and orbital parameters, providing an excellent candidate
for complex terrestrial architectures.

There are some caveats to the analysis presented here that are worth
repeating. First, Section~\ref{smass} describes the use of 3~Gyr
isochrones, which are adopted in order to include stars at the higher
mass end of the stellar mass range. However, the star will evolve with
time, as will the HZ \citep{gallet2017a}, resulting in an increase in
both the width and orbital distance of the HZ. Subsequently, planets
located at the inner edge of the HZ at 3~Gyr will transition into the
Venus Zone \citep{kane2014e} where runaway greenhouse scenarios may
become dominant. Second, our simulations account for Earth-mass
planets and circular orbits. Higher masses and non-zero eccentricities
will reduce orbital stability within the HZ, but lower mass planets
may allow for additional planets, with the exception of strong
resonance regions.

The formation and final architecture of planetary systems depend on
disk mass and migration scenarios
\citep{raymond2005c,morbidelli2016b,dempsey2020a}. Indeed, formation
simulations have demonstrated that terrestrial planets in the HZ are
more likely to occur in the absence of giant planets beyond the snow
line \citep{raymond2012,deelia2013}. Thus, the presence of multiple
terrestrial planets in the HZ of a particular system depends on
numerous factors related to the specific disk properties and formation
processes. However, it should be noted that dynamical packing of
terrestrial planets can have implications for, and even remove, other
planetary and system properties that may be important for
habitability. For example, although compact systems can clearly exist
within the HZ, their potential for harboring moons are considerably
truncated by the reduced Hill radii \citep{kane2017c}. Similarly,
dynamically packed systems around solar-type stars may experience
long-term perturbations that limit the presence of moons for such
planets. It is therefore worth considering in detail how HZ dynamical
packing influences overall planetary evolution.

If such dynamical packing scenarios do exist in relative abundance,
one could speculate as to why such systems have yet to be
discovered. The difficulty with transit detection is imposed by both
the transit probability, the required duration of the survey, and the
assumption of coplanarity with the HZ regime. Likewise,
Table~\ref{signatures} highlights the challenges of RV detection given
the relatively small expected semi-amplitude of the planetary
signatures. In the coming years, the next generation of precision RV
instruments may be able to address this challenge. Current RV surveys
can rule out the presence of giant planets in nearby systems
\citep[e.g.,][and references therein]{fischer2016}, thus
providing potential targets that maximize the likelihood for dynamical
packing. Further into the NASA mission timeline, space-based direct
imaging missions, such as LUVOIR and HabEx, may acquire family
portraits that finally reveal the true HZ architecture of planetary
systems without giant planets.


\section*{Acknowledgements}

The authors would like to thank Paul Dalba, Joshua Pepper, and the
anonymous referee for their constructive feedback on the
manuscript. We also thank Debra Fischer, John Johnson, Kathryn Peek,
and Jason Wright for their contributions to the Keck observations. We
gratefully acknowledge the efforts and dedication of the Keck
Observatory staff for support of HIRES and remote observing. We
recognize and acknowledge the cultural role and reverence that the
summit of Maunakea has within the indigenous Hawaiian community. We
are deeply grateful to have the opportunity to conduct observations
from this mountain. We thank Ken and Gloria Levy, who supported the
construction of the Levy Spectrometer on the Automated Planet
Finder. We thank the University of California and Google for
supporting Lick Observatory and the UCO staff for their dedicated work
scheduling and operating the telescopes of Lick Observatory. This
research has made use of the following archives: the Habitable Zone
Gallery at hzgallery.org and the NASA Exoplanet Archive, which is
operated by the California Institute of Technology, under contract
with the National Aeronautics and Space Administration under the
Exoplanet Exploration Program. The results reported herein benefited
from collaborations and/or information exchange within NASA's Nexus
for Exoplanet System Science (NExSS) research coordination network
sponsored by NASA's Science Mission Directorate.


\software{Mercury \citep{chambers1999}, RadVel \citep{fulton2018a}}



\begin{thebibliography}{}
\expandafter\ifx\csname natexlab\endcsname\relax\def\natexlab#1{#1}\fi

\bibitem[{{Agnew} {et~al.}(2019){Agnew}, {Maddison}, {Horner}, \&
  {Kane}}]{agnew2019}
{Agnew}, M.~T., {Maddison}, S.~T., {Horner}, J., \& {Kane}, S.~R. 2019, \mnras,
  485, 4703

\bibitem[{{Akeson} {et~al.}(2013){Akeson}, {Chen}, {Ciardi}, {Crane}, {Good},
  {Harbut}, {Jackson}, {Kane}, {Laity}, {Leifer}, {Lynn}, {McElroy}, {Papin},
  {Plavchan}, {Ram{\'\i}rez}, {Rey}, {von Braun}, {Wittman}, {Abajian}, {Ali},
  {Beichman}, {Beekley}, {Berriman}, {Berukoff}, {Bryden}, {Chan}, {Groom},
  {Lau}, {Payne}, {Regelson}, {Saucedo}, {Schmitz}, {Stauffer}, {Wyatt}, \&
  {Zhang}}]{akeson2013}
{Akeson}, R.~L., {Chen}, X., {Ciardi}, D., {et~al.} 2013, \pasp, 125, 989

\bibitem[{{Anglada-Escud{\'e}} {et~al.}(2013){Anglada-Escud{\'e}}, {Tuomi},
  {Gerlach}, {Barnes}, {Heller}, {Jenkins}, {Wende}, {Vogt}, {Butler},
  {Reiners}, \& {Jones}}]{angladaescude2013c}
{Anglada-Escud{\'e}}, G., {Tuomi}, M., {Gerlach}, E., {et~al.} 2013, \aap, 556,
  A126

\bibitem[{{Arney}(2019)}]{arney2019}
{Arney}, G.~N. 2019, \apjl, 873, L7

\bibitem[{{Barnes} \& {Quinn}(2004)}]{barnes2004a}
{Barnes}, R., \& {Quinn}, T. 2004, \apj, 611, 494

\bibitem[{{Batygin} \& {Laughlin}(2015)}]{batygin2015b}
{Batygin}, K., \& {Laughlin}, G. 2015, Proceedings of the National Academy of
  Science, 112, 4214

\bibitem[{{Becker} \& {Adams}(2017)}]{becker2017a}
{Becker}, J.~C., \& {Adams}, F.~C. 2017, \mnras, 468, 549

\bibitem[{{Bonfils} {et~al.}(2013){Bonfils}, {Delfosse}, {Udry}, {Forveille},
  {Mayor}, {Perrier}, {Bouchy}, {Gillon}, {Lovis}, {Pepe}, {Queloz}, {Santos},
  {S{\'e}gransan}, \& {Bertaux}}]{bonfils2013a}
{Bonfils}, X., {Delfosse}, X., {Udry}, S., {et~al.} 2013, \aap, 549, A109

\bibitem[{{Boyajian} {et~al.}(2012){Boyajian}, {McAlister}, {van Belle},
  {Gies}, {ten Brummelaar}, {von Braun}, {Farrington}, {Goldfinger}, {O'Brien},
  {Parks}, {Richardson}, {Ridgway}, {Schaefer}, {Sturmann}, {Sturmann},
  {Touhami}, {Turner}, \& {White}}]{boyajian2012a}
{Boyajian}, T.~S., {McAlister}, H.~A., {van Belle}, G., {et~al.} 2012, \apj,
  746, 101

\bibitem[{{Brasser} {et~al.}(2009){Brasser}, {Morbidelli}, {Gomes}, {Tsiganis},
  \& {Levison}}]{brasser2009}
{Brasser}, R., {Morbidelli}, A., {Gomes}, R., {Tsiganis}, K., \& {Levison},
  H.~F. 2009, \aap, 507, 1053

\bibitem[{{Catanzarite} \& {Shao}(2011)}]{catanzarite2011a}
{Catanzarite}, J., \& {Shao}, M. 2011, \pasp, 123, 171

\bibitem[{{Chambers}(1999)}]{chambers1999}
{Chambers}, J.~E. 1999, \mnras, 304, 793

\bibitem[{{Choi} {et~al.}(2016){Choi}, {Dotter}, {Conroy}, {Cantiello},
  {Paxton}, \& {Johnson}}]{choi2016}
{Choi}, J., {Dotter}, A., {Conroy}, C., {et~al.} 2016, \apj, 823, 102

\bibitem[{{Cumming} {et~al.}(2008){Cumming}, {Butler}, {Marcy}, {Vogt},
  {Wright}, \& {Fischer}}]{cumming2008}
{Cumming}, A., {Butler}, R.~P., {Marcy}, G.~W., {et~al.} 2008, \pasp, 120, 531

\bibitem[{{de El{\'\i}a} {et~al.}(2013){de El{\'\i}a}, {Guilera}, \&
  {Brunini}}]{deelia2013}
{de El{\'\i}a}, G.~C., {Guilera}, O.~M., \& {Brunini}, A. 2013, \aap, 557, A42

\bibitem[{{Dempsey} {et~al.}(2020){Dempsey}, {Lee}, \&
  {Lithwick}}]{dempsey2020a}
{Dempsey}, A.~M., {Lee}, W.-K., \& {Lithwick}, Y. 2020, \apj, 891, 108

\bibitem[{{Dotter}(2016)}]{dotter2016}
{Dotter}, A. 2016, \apjs, 222, 8

\bibitem[{{Dressing} \& {Charbonneau}(2013)}]{dressing2013}
{Dressing}, C.~D., \& {Charbonneau}, D. 2013, \apj, 767, 95

\bibitem[{{Dressing} \& {Charbonneau}(2015)}]{dressing2015b}
---. 2015, \apj, 807, 45

\bibitem[{{Dulz} {et~al.}(2020){Dulz}, {Plavchan}, {Crepp}, {Stark}, {Morgan},
  {Kane}, {Newman}, {Matzko}, \& {Mulders}}]{dulz2020}
{Dulz}, S.~D., {Plavchan}, P., {Crepp}, J.~R., {et~al.} 2020, \apj, 893, 122

\bibitem[{{Duncan} {et~al.}(1998){Duncan}, {Levison}, \& {Lee}}]{duncan1998}
{Duncan}, M.~J., {Levison}, H.~F., \& {Lee}, M.~H. 1998, \aj, 116, 2067

\bibitem[{{Endl} {et~al.}(2006){Endl}, {Cochran}, {K{\"u}rster}, {Paulson},
  {Wittenmyer}, {MacQueen}, \& {Tull}}]{endl2006b}
{Endl}, M., {Cochran}, W.~D., {K{\"u}rster}, M., {et~al.} 2006, \apj, 649, 436

\bibitem[{{Fang} \& {Margot}(2013)}]{fang2013}
{Fang}, J., \& {Margot}, J.-L. 2013, \apj, 767, 115

\bibitem[{{Fischer} {et~al.}(2016){Fischer}, {Anglada-Escude}, {Arriagada},
  {Baluev}, {Bean}, {Bouchy}, {Buchhave}, {Carroll}, {Chakraborty}, {Crepp},
  {Dawson}, {Diddams}, {Dumusque}, {Eastman}, {Endl}, {Figueira}, {Ford},
  {Foreman-Mackey}, {Fournier}, {F{\H{u}}r{\'e}sz}, {Gaudi}, {Gregory},
  {Grundahl}, {Hatzes}, {H{\'e}brard}, {Herrero}, {Hogg}, {Howard}, {Johnson},
  {Jorden}, {Jurgenson}, {Latham}, {Laughlin}, {Loredo}, {Lovis}, {Mahadevan},
  {McCracken}, {Pepe}, {Perez}, {Phillips}, {Plavchan}, {Prato}, {Quirrenbach},
  {Reiners}, {Robertson}, {Santos}, {Sawyer}, {Segransan}, {Sozzetti},
  {Steinmetz}, {Szentgyorgyi}, {Udry}, {Valenti}, {Wang}, {Wittenmyer}, \&
  {Wright}}]{fischer2016}
{Fischer}, D.~A., {Anglada-Escude}, G., {Arriagada}, P., {et~al.} 2016, \pasp,
  128, 066001

\bibitem[{{Ford}(2014)}]{ford2014}
{Ford}, E.~B. 2014, Proceedings of the National Academy of Science, 111, 12616

\bibitem[{{Foreman-Mackey} {et~al.}(2014){Foreman-Mackey}, {Hogg}, \&
  {Morton}}]{foremanmackey2014}
{Foreman-Mackey}, D., {Hogg}, D.~W., \& {Morton}, T.~D. 2014, \apj, 795, 64

\bibitem[{{Fulton} {et~al.}(2018){Fulton}, {Petigura}, {Blunt}, \&
  {Sinukoff}}]{fulton2018a}
{Fulton}, B.~J., {Petigura}, E.~A., {Blunt}, S., \& {Sinukoff}, E. 2018, \pasp,
  130, 044504

\bibitem[{{Gaidos}(2013)}]{gaidos2013b}
{Gaidos}, E. 2013, \apj, 770, 90

\bibitem[{{Gallet} {et~al.}(2017){Gallet}, {Charbonnel}, {Amard}, {Brun},
  {Palacios}, \& {Mathis}}]{gallet2017a}
{Gallet}, F., {Charbonnel}, C., {Amard}, L., {et~al.} 2017, \aap, 597, A14

\bibitem[{{Gaudi} {et~al.}(2020){Gaudi}, {Seager}, {Mennesson}, {Kiessling},
  {Warfield}, {Cahoy}, {Clarke}, {Domagal-Goldman}, {Feinberg}, {Guyon},
  {Kasdin}, {Mawet}, {Plavchan}, {Robinson}, {Rogers}, {Scowen}, {Somerville},
  {Stapelfeldt}, {Stark}, {Stern}, {Turnbull}, {Amini}, {Kuan}, {Martin},
  {Morgan}, {Redding}, {Stahl}, {Webb}, {Alvarez-Salazar}, {Arnold}, {Arya},
  {Balasubramanian}, {Baysinger}, {Bell}, {Below}, {Benson}, {Blais}, {Booth},
  {Bourgeois}, {Bradford}, {Brewer}, {Brooks}, {Cady}, {Caldwell}, {Calvet},
  {Carr}, {Chan}, {Cormarkovic}, {Coste}, {Cox}, {Danner}, {Davis}, {Dewell},
  {Dorsett}, {Dunn}, {East}, {Effinger}, {Eng}, {Freebury}, {Garcia}, {Gaskin},
  {Greene}, {Hennessy}, {Hilgemann}, {Hood}, {Holota}, {Howe}, {Huang}, {Hull},
  {Hunt}, {Hurd}, {Johnson}, {Kissil}, {Knight}, {Kolenz}, {Kraus}, {Krist},
  {Li}, {Lisman}, {Mandic}, {Mann}, {Marchen}, {Marrese-Reading}, {McCready},
  {McGown}, {Missun}, {Miyaguchi}, {Moore}, {Nemati}, {Nikzad}, {Nissen},
  {Novicki}, {Perrine}, {Pineda}, {Polanco}, {Putnam}, {Qureshi}, {Richards},
  {Eldorado Riggs}, {Rodgers}, {Rud}, {Saini}, {Scalisi}, {Scharf}, {Schulz},
  {Serabyn}, {Sigrist}, {Sikkia}, {Singleton}, {Shaklan}, {Smith}, {Southerd},
  {Stahl}, {Steeves}, {Sturges}, {Sullivan}, {Tang}, {Taras}, {Tesch},
  {Therrell}, {Tseng}, {Valente}, {Van Buren}, {Villalvazo}, {Warwick}, {Webb},
  {Westerhoff}, {Wofford}, {Wu}, {Woo}, {Wood}, {Ziemer}, {Arney}, {Anderson},
  {Ma{\'\i}z-Apell{\'a}niz}, {Bartlett}, {Belikov}, {Bendek}, {Cenko},
  {Douglas}, {Dulz}, {Evans}, {Faramaz}, {Feng}, {Ferguson}, {Follette},
  {Ford}, {Garc{\'\i}a}, {Geha}, {Gelino}, {G{\"o}tberg}, {Hildebrand t}, {Hu},
  {Jahnke}, {Kennedy}, {Kreidberg}, {Isella}, {Lopez}, {Marchis}, {Macri},
  {Marley}, {Matzko}, {Mazoyer}, {McCandliss}, {Meshkat}, {Mordasini},
  {Morris}, {Nielsen}, {Newman}, {Petigura}, {Postman}, {Reines}, {Roberge},
  {Roederer}, {Ruane}, {Schwieterman}, {Sirbu}, {Spalding}, {Teplitz},
  {Tumlinson}, {Turner}, {Werk}, {Wofford}, {Wyatt}, {Young}, \&
  {Zellem}}]{reporthabex}
{Gaudi}, B.~S., {Seager}, S., {Mennesson}, B., {et~al.} 2020, arXiv e-prints,
  arXiv:2001.06683

\bibitem[{{Gillon} {et~al.}(2017){Gillon}, {Triaud}, {Demory}, {Jehin}, {Agol},
  {Deck}, {Lederer}, {de Wit}, {Burdanov}, {Ingalls}, {Bolmont}, {Leconte},
  {Raymond}, {Selsis}, {Turbet}, {Barkaoui}, {Burgasser}, {Burleigh}, {Carey},
  {Chaushev}, {Copperwheat}, {Delrez}, {Fernandes}, {Holdsworth}, {Kotze}, {Van
  Grootel}, {Almleaky}, {Benkhaldoun}, {Magain}, \& {Queloz}}]{gillon2017a}
{Gillon}, M., {Triaud}, A.~H.~M.~J., {Demory}, B.-O., {et~al.} 2017, \nat, 542,
  456

\bibitem[{{Henry}(1999)}]{henry1999}
{Henry}, G.~W. 1999, \pasp, 111, 845

\bibitem[{{Herrero} {et~al.}(2012){Herrero}, {Ribas}, {Jordi}, {Guinan}, \&
  {Engle}}]{herrero2012}
{Herrero}, E., {Ribas}, I., {Jordi}, C., {Guinan}, E.~F., \& {Engle}, S.~G.
  2012, \aap, 537, A147

\bibitem[{{Hill} {et~al.}(2018){Hill}, {Kane}, {Seperuelo Duarte}, {Kopparapu},
  {Gelino}, \& {Wittenmyer}}]{hill2018}
{Hill}, M.~L., {Kane}, S.~R., {Seperuelo Duarte}, E., {et~al.} 2018, \apj, 860,
  67

\bibitem[{{Horner} \& {Jones}(2008)}]{horner2008a}
{Horner}, J., \& {Jones}, B.~W. 2008, International Journal of Astrobiology, 7,
  251

\bibitem[{{Horner} {et~al.}(2020){Horner}, {Vervoort}, {Kane}, {Ceja},
  {Waltham}, {Gilmore}, \& {Kirtland Turner}}]{horner2020a}
{Horner}, J., {Vervoort}, P., {Kane}, S.~R., {et~al.} 2020, \aj, 159, 10

\bibitem[{{Howard} \& {Fulton}(2016)}]{howard2016}
{Howard}, A.~W., \& {Fulton}, B.~J. 2016, \pasp, 128, 114401

\bibitem[{{Johns} {et~al.}(2018){Johns}, {Marti}, {Huff}, {McCann},
  {Wittenmyer}, {Horner}, \& {Wright}}]{johns2018}
{Johns}, D., {Marti}, C., {Huff}, M., {et~al.} 2018, \apjs, 239, 14

\bibitem[{{Johnson} {et~al.}(2010){Johnson}, {Aller}, {Howard}, \&
  {Crepp}}]{johnson2010d}
{Johnson}, J.~A., {Aller}, K.~M., {Howard}, A.~W., \& {Crepp}, J.~R. 2010,
  \pasp, 122, 905

\bibitem[{{Kane}(2015)}]{kane2015b}
{Kane}, S.~R. 2015, \apjl, 814, L9

\bibitem[{{Kane}(2017)}]{kane2017c}
---. 2017, \apjl, 839, L19

\bibitem[{{Kane}(2018)}]{kane2018a}
---. 2018, \apjl, 861, L21

\bibitem[{{Kane} \& {Blunt}(2019)}]{kane2019e}
{Kane}, S.~R., \& {Blunt}, S. 2019, \aj, 158, 209

\bibitem[{{Kane} \& {Gelino}(2012)}]{kane2012a}
{Kane}, S.~R., \& {Gelino}, D.~M. 2012, \pasp, 124, 323

\bibitem[{{Kane} {et~al.}(2013){Kane}, {Hinkel}, \& {Raymond}}]{kane2013e}
{Kane}, S.~R., {Hinkel}, N.~R., \& {Raymond}, S.~N. 2013, \aj, 146, 122

\bibitem[{{Kane} {et~al.}(2014){Kane}, {Kopparapu}, \&
  {Domagal-Goldman}}]{kane2014e}
{Kane}, S.~R., {Kopparapu}, R.~K., \& {Domagal-Goldman}, S.~D. 2014, \apjl,
  794, L5

\bibitem[{{Kane} {et~al.}(2018){Kane}, {Meshkat}, \& {Turnbull}}]{kane2018c}
{Kane}, S.~R., {Meshkat}, T., \& {Turnbull}, M.~C. 2018, \aj, 156, 267

\bibitem[{{Kane} \& {von Braun}(2008)}]{kane2008b}
{Kane}, S.~R., \& {von Braun}, K. 2008, \apj, 689, 492

\bibitem[{{Kane} {et~al.}(2016){Kane}, {Hill}, {Kasting}, {Kopparapu},
  {Quintana}, {Barclay}, {Batalha}, {Borucki}, {Ciardi}, {Haghighipour},
  {Hinkel}, {Kaltenegger}, {Selsis}, \& {Torres}}]{kane2016c}
{Kane}, S.~R., {Hill}, M.~L., {Kasting}, J.~F., {et~al.} 2016, \apj, 830, 1

\bibitem[{{Kasting} {et~al.}(1993){Kasting}, {Whitmire}, \&
  {Reynolds}}]{kasting1993a}
{Kasting}, J.~F., {Whitmire}, D.~P., \& {Reynolds}, R.~T. 1993, \icarus, 101,
  108

\bibitem[{{Konacki}(2005)}]{konacki2005b}
{Konacki}, M. 2005, \apj, 626, 431

\bibitem[{{Kopparapu}(2013)}]{kopparapu2013b}
{Kopparapu}, R.~K. 2013, \apj, 767, L8

\bibitem[{{Kopparapu} \& {Barnes}(2010)}]{kopparapu2010}
{Kopparapu}, R.~K., \& {Barnes}, R. 2010, \apj, 716, 1336

\bibitem[{{Kopparapu} {et~al.}(2014){Kopparapu}, {Ramirez}, {SchottelKotte},
  {Kasting}, {Domagal-Goldman}, \& {Eymet}}]{kopparapu2014}
{Kopparapu}, R.~K., {Ramirez}, R.~M., {SchottelKotte}, J., {et~al.} 2014, \apj,
  787, L29

\bibitem[{{Kopparapu} {et~al.}(2009){Kopparapu}, {Raymond}, \&
  {Barnes}}]{kopparapu2009}
{Kopparapu}, R.~K., {Raymond}, S.~N., \& {Barnes}, R. 2009, \apj, 695, L181

\bibitem[{{Kopparapu} {et~al.}(2013){Kopparapu}, {Ramirez}, {Kasting}, {Eymet},
  {Robinson}, {Mahadevan}, {Terrien}, {Domagal-Goldman}, {Meadows}, \&
  {Deshpande}}]{kopparapu2013a}
{Kopparapu}, R.~K., {Ramirez}, R., {Kasting}, J.~F., {et~al.} 2013, \apj, 765,
  131

\bibitem[{{Laskar}(1996)}]{laskar1996b}
{Laskar}, J. 1996, Celestial Mechanics and Dynamical Astronomy, 64, 115

\bibitem[{{Lecar} {et~al.}(2001){Lecar}, {Franklin}, {Holman}, \&
  {Murray}}]{lecar2001}
{Lecar}, M., {Franklin}, F.~A., {Holman}, M.~J., \& {Murray}, N.~J. 2001,
  \araa, 39, 581

\bibitem[{{Levison} \& {Agnor}(2003)}]{levison2003a}
{Levison}, H.~F., \& {Agnor}, C. 2003, \aj, 125, 2692

\bibitem[{{Lissauer} {et~al.}(2001){Lissauer}, {Quintana}, {Rivera}, \&
  {Duncan}}]{lissauer2001c}
{Lissauer}, J.~J., {Quintana}, E.~V., {Rivera}, E.~J., \& {Duncan}, M.~J. 2001,
  \icarus, 154, 449

\bibitem[{{Luger} {et~al.}(2017){Luger}, {Sestovic}, {Kruse}, {Grimm},
  {Demory}, {Agol}, {Bolmont}, {Fabrycky}, {Fernandes}, {Van Grootel},
  {Burgasser}, {Gillon}, {Ingalls}, {Jehin}, {Raymond}, {Selsis}, {Triaud},
  {Barclay}, {Barentsen}, {Howell}, {Delrez}, {de Wit}, {Foreman-Mackey},
  {Holdsworth}, {Leconte}, {Lederer}, {Turbet}, {Almleaky}, {Benkhaldoun},
  {Magain}, {Morris}, {Heng}, \& {Queloz}}]{luger2017b}
{Luger}, R., {Sestovic}, M., {Kruse}, E., {et~al.} 2017, Nature Astronomy, 1,
  0129

\bibitem[{{Matsumura} {et~al.}(2013){Matsumura}, {Ida}, \&
  {Nagasawa}}]{matsumura2013}
{Matsumura}, S., {Ida}, S., \& {Nagasawa}, M. 2013, \apj, 767, 129

\bibitem[{{Morbidelli} \& {Raymond}(2016)}]{morbidelli2016b}
{Morbidelli}, A., \& {Raymond}, S.~N. 2016, Journal of Geophysical Research
  (Planets), 121, 1962

\bibitem[{{Murray} \& {Holman}(2001)}]{murray2001a}
{Murray}, N., \& {Holman}, M. 2001, \nat, 410, 773

\bibitem[{{Obertas} {et~al.}(2017){Obertas}, {Van Laerhoven}, \&
  {Tamayo}}]{obertas2017}
{Obertas}, A., {Van Laerhoven}, C., \& {Tamayo}, D. 2017, \icarus, 293, 52

\bibitem[{{Porto de Mello} {et~al.}(2006){Porto de Mello}, {del Peloso}, \&
  {Ghezzi}}]{portodemello2006}
{Porto de Mello}, G., {del Peloso}, E.~F., \& {Ghezzi}, L. 2006, Astrobiology,
  6, 308

\bibitem[{{Pu} \& {Wu}(2015)}]{pu2015}
{Pu}, B., \& {Wu}, Y. 2015, \apj, 807, 44

\bibitem[{{Radick} {et~al.}(2018){Radick}, {Lockwood}, {Henry}, {Hall}, \&
  {Pevtsov}}]{radick2018}
{Radick}, R.~R., {Lockwood}, G.~W., {Henry}, G.~W., {Hall}, J.~C., \&
  {Pevtsov}, A.~A. 2018, \apj, 855, 75

\bibitem[{{Radovan} {et~al.}(2014){Radovan}, {Lanclos}, {Holden}, {Kibrick},
  {Allen}, {Deich}, {Rivera}, {Burt}, {Fulton}, {Butler}, \&
  {Vogt}}]{radovan2014}
{Radovan}, M.~V., {Lanclos}, K., {Holden}, B.~P., {et~al.} 2014, Society of
  Photo-Optical Instrumentation Engineers (SPIE) Conference Series, Vol. 9145,
  {The automated planet finder at Lick Observatory} (SPIE Press), 91452B

\bibitem[{{Raymond} {et~al.}(2005){Raymond}, {Quinn}, \&
  {Lunine}}]{raymond2005c}
{Raymond}, S.~N., {Quinn}, T., \& {Lunine}, J.~I. 2005, \apj, 632, 670

\bibitem[{{Raymond} {et~al.}(2012){Raymond}, {Armitage}, {Moro-Mart{\'\i}n},
  {Booth}, {Wyatt}, {Armstrong}, {Mand ell}, {Selsis}, \& {West}}]{raymond2012}
{Raymond}, S.~N., {Armitage}, P.~J., {Moro-Mart{\'\i}n}, A., {et~al.} 2012,
  \aap, 541, A11

\bibitem[{{Shields} {et~al.}(2016){Shields}, {Ballard}, \&
  {Johnson}}]{shields2016b}
{Shields}, A.~L., {Ballard}, S., \& {Johnson}, J.~A. 2016, \physrep, 663, 1

\bibitem[{{Smith} \& {Lissauer}(2009)}]{smith2009}
{Smith}, A.~W., \& {Lissauer}, J.~J. 2009, \icarus, 201, 381

\bibitem[{{Smith} \& {Lissauer}(2010)}]{smith2010}
---. 2010, Celestial Mechanics and Dynamical Astronomy, 107, 487

\bibitem[{{The LUVOIR Team}(2019)}]{reportluvoir}
{The LUVOIR Team}. 2019, arXiv e-prints, arXiv:1912.06219

\bibitem[{{Tremaine}(2015)}]{tremaine2015}
{Tremaine}, S. 2015, \apj, 807, 157

\bibitem[{{Tsiganis} {et~al.}(2005){Tsiganis}, {Gomes}, {Morbidelli}, \&
  {Levison}}]{tsiganis2005b}
{Tsiganis}, K., {Gomes}, R., {Morbidelli}, A., \& {Levison}, H.~F. 2005, \nat,
  435, 459

\bibitem[{{Turnbull}(2015)}]{turnbull2015}
{Turnbull}, M.~C. 2015, arXiv e-prints, arXiv:1510.01731

\bibitem[{{Valenti} \& {Fischer}(2005)}]{valenti2005}
{Valenti}, J.~A., \& {Fischer}, D.~A. 2005, \apjs, 159, 141

\bibitem[{{van Belle} \& {von Braun}(2009)}]{vanbelle2009a}
{van Belle}, G.~T., \& {von Braun}, K. 2009, \apj, 694, 1085

\bibitem[{{Vogt} {et~al.}(1994){Vogt}, {Allen}, {Bigelow}, {Bresee}, {Brown},
  {Cantrall}, {Conrad}, {Couture}, {Delaney}, {Epps}, {Hilyard}, {Hilyard},
  {Horn}, {Jern}, {Kanto}, {Keane}, {Kibrick}, {Lewis}, {Osborne},
  {Pardeilhan}, {Pfister}, {Ricketts}, {Robinson}, {Stover}, {Tucker}, {Ward},
  \& {Wei}}]{vogt1994}
{Vogt}, S.~S., {Allen}, S.~L., {Bigelow}, B.~C., {et~al.} 1994, Society of
  Photo-Optical Instrumentation Engineers (SPIE) Conference Series, Vol. 2198,
  {HIRES: the high-resolution echelle spectrometer on the Keck 10-m Telescope}
  (SPIE Press), 362

\bibitem[{{Vogt} {et~al.}(2014){Vogt}, {Radovan}, {Kibrick}, {Butler},
  {Alcott}, {Allen}, {Arriagada}, {Bolte}, {Burt}, {Cabak}, {Chloros},
  {Cowley}, {Deich}, {Dupraw}, {Earthman}, {Epps}, {Faber}, {Fischer}, {Gates},
  {Hilyard}, {Holden}, {Johnston}, {Keiser}, {Kanto}, {Katsuki}, {Laiterman},
  {Lanclos}, {Laughlin}, {Lewis}, {Lockwood}, {Lynam}, {Marcy}, {McLean},
  {Miller}, {Misch}, {Peck}, {Pfister}, {Phillips}, {Rivera}, {Sand ford},
  {Saylor}, {Stover}, {Thompson}, {Walp}, {Ward}, {Wareham}, {Wei}, \&
  {Wright}}]{vogt2014a}
{Vogt}, S.~S., {Radovan}, M., {Kibrick}, R., {et~al.} 2014, \pasp, 126, 359

\bibitem[{{Weiss} {et~al.}(2018){Weiss}, {Marcy}, {Petigura}, {Fulton},
  {Howard}, {Winn}, {Isaacson}, {Morton}, {Hirsch}, {Sinukoff}, {Cumming},
  {Hebb}, \& {Cargile}}]{weiss2018a}
{Weiss}, L.~M., {Marcy}, G.~W., {Petigura}, E.~A., {et~al.} 2018, \aj, 155, 48

\bibitem[{{Wisdom}(1983)}]{wisdom1983a}
{Wisdom}, J. 1983, \icarus, 56, 51

\bibitem[{{Wisdom}(2006)}]{wisdom2006b}
---. 2006, \aj, 131, 2294

\bibitem[{{Wisdom} \& {Holman}(1991)}]{wisdom1991}
{Wisdom}, J., \& {Holman}, M. 1991, \aj, 102, 1528

\bibitem[{{Wittenmyer} {et~al.}(2016){Wittenmyer}, {Butler}, {Tinney},
  {Horner}, {Carter}, {Wright}, {Jones}, {Bailey}, \&
  {O'Toole}}]{wittenmyer2016c}
{Wittenmyer}, R.~A., {Butler}, R.~P., {Tinney}, C.~G., {et~al.} 2016, \apj,
  819, 28

\bibitem[{{Wittenmyer} {et~al.}(2020){Wittenmyer}, {Butler}, {Horner}, {Clark},
  {Tinney}, {Carter}, {Wang}, {Johnson}, \& {Collins}}]{wittenmyer2020a}
{Wittenmyer}, R.~A., {Butler}, R.~P., {Horner}, J., {et~al.} 2020, \mnras, 491,
  5248

\bibitem[{{Zechmeister} {et~al.}(2013){Zechmeister}, {K{\"u}rster}, {Endl}, {Lo
  Curto}, {Hartman}, {Nilsson}, {Henning}, {Hatzes}, \&
  {Cochran}}]{zechmeister2013}
{Zechmeister}, M., {K{\"u}rster}, M., {Endl}, M., {et~al.} 2013, \aap, 552, A78

\bibitem[{{Zechmeister} {et~al.}(2019){Zechmeister}, {Dreizler}, {Ribas},
  {Reiners}, {Caballero}, {Bauer}, {B{\'e}jar}, {Gonz{\'a}lez-Cuesta},
  {Herrero}, {Lalitha}, {L{\'o}pez-Gonz{\'a}lez}, {Luque}, {Morales},
  {Pall{\'e}}, {Rodr{\'\i}guez}, {Rodr{\'\i}guez L{\'o}pez}, {Tal-Or},
  {Anglada-Escud{\'e}}, {Quirrenbach}, {Amado}, {Abril}, {Aceituno},
  {Aceituno}, {Alonso-Floriano}, {Ammler-von Eiff}, {Antona Jim{\'e}nez},
  {Anwand-Heerwart}, {Arroyo-Torres}, {Azzaro}, {Baroch}, {Barrado},
  {Becerril}, {Ben{\'\i}tez}, {Berdi{\~n}as}, {Bergond}, {Bluhm},
  {Brinkm{\"o}ller}, {del Burgo}, {Calvo Ortega}, {Cano}, {Cardona
  Guill{\'e}n}, {Carro}, {C{\'a}rdenas V{\'a}zquez}, {Casal},
  {Casasayas-Barris}, {Casanova}, {Chaturvedi}, {Cifuentes}, {Claret},
  {Colom{\'e}}, {Cort{\'e}s-Contreras}, {Czesla}, {D{\'\i}ez-Alonso}, {Dorda},
  {Fern{\'a}ndez}, {Fern{\'a}ndez-Mart{\'\i}n}, {Fuhrmeister}, {Fukui},
  {Galad{\'\i}-Enr{\'\i}quez}, {Gallardo Cava}, {Garcia de la Fuente},
  {Garcia-Piquer}, {Garc{\'\i}a Vargas}, {Gesa}, {G{\'o}ngora Rueda},
  {Gonz{\'a}lez-{\'A}lvarez}, {Gonz{\'a}lez Hern{\'a}ndez},
  {Gonz{\'a}lez-Peinado}, {Gr{\"o}zinger}, {Gu{\`a}rdia}, {Guijarro}, {de
  Guindos}, {Hatzes}, {Hauschildt}, {Hedrosa}, {Helmling}, {Henning},
  {Hermelo}, {Hern{\'a}ndez Arabi}, {Hern{\'a}ndez Casta{\~n}o}, {Hern{\'a}ndez
  Otero}, {Hintz}, {Huke}, {Huber}, {Jeffers}, {Johnson}, {de Juan},
  {Kaminski}, {Kemmer}, {Kim}, {Klahr}, {Klein}, {Kl{\"u}ter}, {Klutsch},
  {Kossakowski}, {K{\"u}rster}, {Labarga}, {Lafarga}, {Llamas}, {Lamp{\'o}n},
  {Lara}, {Launhardt}, {L{\'a}zaro}, {Lodieu}, {L{\'o}pez del Fresno},
  {L{\'o}pez-Puertas}, {L{\'o}pez Salas}, {L{\'o}pez-Santiago}, {Mag{\'a}n
  Madinabeitia}, {Mall}, {Mancini}, {Mand el}, {Marfil}, {Mar{\'\i}n Molina},
  {Maroto Fern{\'a}ndez}, {Mart{\'\i}n}, {Mart{\'\i}n-Fern{\'a}ndez},
  {Mart{\'\i}n-Ruiz}, {Marvin}, {Mirabet}, {Monta{\~n}{\'e}s-Rodr{\'\i}guez},
  {Montes}, {Moreno-Raya}, {Nagel}, {Naranjo}, {Narita}, {Nortmann}, {Nowak},
  {Ofir}, {Oshagh}, {Panduro}, {Parviainen}, {Pascual}, {Passegger}, {Pavlov},
  {Pedraz}, {P{\'e}rez-Calpena}, {P{\'e}rez Medialdea}, {Perger}, {Perryman},
  {Rabaza}, {Ram{\'o}n Ballesta}, {Rebolo}, {Redondo}, {Reffert}, {Reinhardt},
  {Rhode}, {Rix}, {Rodler}, {Rodr{\'\i}guez Trinidad}, {Rosich}, {Sadegi},
  {S{\'a}nchez-Blanco}, {S{\'a}nchez Carrasco}, {S{\'a}nchez-L{\'o}pez},
  {Sanz-Forcada}, {Sarkis}, {Sarmiento}, {Sch{\"a}fer}, {Schmitt},
  {Sch{\"o}fer}, {Schweitzer}, {Seifert}, {Shulyak}, {Solano}, {Sota}, {Stahl},
  {Stock}, {Strachan}, {Stuber}, {St{\"u}rmer}, {Su{\'a}rez}, {Tabernero},
  {Tala Pinto}, {Trifonov}, {Veredas}, {Vico Linares}, {Vilardell}, {Wagner},
  {Wolthoff}, {Xu}, {Yan}, \& {Zapatero Osorio}}]{zechmeister2019}
{Zechmeister}, M., {Dreizler}, S., {Ribas}, I., {et~al.} 2019, \aap, 627, A49

\end{thebibliography}


\end{document}